\documentclass[epj,final]{svjour}
\usepackage{graphics}
\usepackage{psfig}
\usepackage{amssymb}

\begin{document}
\title{Convection in horizontally shaken granular material}
\author{Clara Salue\~na\inst{1,2} \and Thorsten P\"oschel\inst{2}}

\institute{
\inst{1} Departament de F\'{\i}sica Fonamental, Divisi\'o de Ci\`encies Experimentals i Matem\`atiques, Universitat de Barcelona, Spain\\
\inst{2} Humboldt-Universit\"at zu Berlin, 
Institut f\"ur Physik, Invalidenstra\ss e 110, D-10115 Berlin, Germany.\\ 
http://summa.physik.hu-berlin.de/$\sim$kies/
}
\date{Received:  / Revised version: }

\abstract{
  In horizontally shaken granular material different types of pattern
  formation have been reported. We want to deal with the convection
  instability which has been observed in experiments and which
  recently has been investigated numerically. Using two dimensional
  molecular dynamics we show that the convection pattern depends
  crucially on the inelastic properties of the material. The concept of
  restitution coefficient provides arguments for the change of the
  behaviour with varying inelasticity.
\PACS{
{81.05.Rm}{Porous materials; granular materials}\and
{83.70.Fn}{Granular solids}\and
{30.My}{Vibrations, aeroelasticity, hydroelasticity, mechanical waves, and shocks}
}
\keywords{Granular convection, horizontal shaking, molecular dynamics}}
\maketitle

\section{Introduction}
When granular material in a cubic container is shaken
horizontally one observes experimentally different types of
instabilities, i.e. spontaneous formation of ripples in shallow
beds~\cite{StrassburgerBetatSchererRehberg:1996},
liquefaction~\cite{RistowStrassburgerRehberg:1997,Ristow:1997}, convective
motion~\cite{TennakoonBehringer:1997,Jaeger} and recurrent swelling of
shaken material where the period of swelling decouples from the
forcing period~\cite{RosenkranzPoeschel:1996}. Other interesting experimental results concerning simultaneously vertically and horizontally vibrated granular systems~\cite{TennakoonBehringer:1998} and enhanced packing of spheres due to horizontal vibrations~\cite{PouliquenNicolasWeidman:1997} have been reported recently. Horizontally shaken
granular systems have been simulated numerically using cellular
automata~\cite{StrassburgerBetatSchererRehberg:1996} as well as
molecular dynamics
techniques~\cite{RistowStrassburgerRehberg:1997,Ristow:1997,IwashitaEtAl:1988,LiffmanMetcalfeCleary:1997,SaluenaEsipovPoeschel:1997,SPEpre99}.
Theoretical work on horizontal shaking can be found
in~\cite{SaluenaEsipovPoeschel:1997} and the dynamics of a single
particle in a horizontally shaken box has been discussed
in~\cite{DrosselPrellberg:1997}.

\begin{figure}[htbp]
 \centerline{\psfig{file=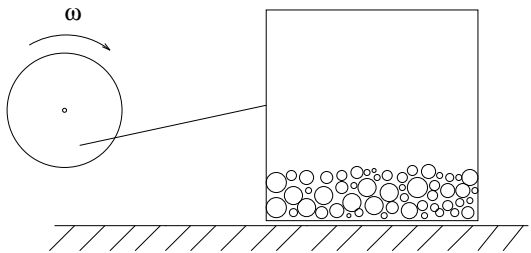,width=7cm,clip=}}  
  \caption{Sketch of the simulated system.}
  \label{fig:sketch}
\end{figure}

Recently the effect of convection in a horizontally shaken box filled with 
granular material attracted much attention and presently the effect is studied
experimentally by different
groups~\cite{TennakoonBehringer:1997,Jaeger,RosenkranzPoeschel:1996}.
Unlike the effect of convective motion in vertically shaken granular
material which has been studied intensively experimentally,
analytically and by means of computer simulations
(s.~e.g.~\cite{vertikalEX,JaegerVert,vertikalANA,vertikalMD}), there
exist only a few references on horizontal shaking. Different from the
vertical case, where the ``architecture'' of the convection pattern is
very simple~\cite{BizonEtAl:1998}, in horizontally shaken containers one observes a variety
of different patterns, convecting in different directions, in parallel
as well as perpendicular to the direction of
forcing~\cite{TennakoonBehringer:1997}. Under certain conditions one
observes several convection rolls on top of each other~\cite{Jaeger}.
An impression of the complicated convection can be found in the
internet~\cite{movies}.

Whereas the properties of convection in vertically sha\-ken systems
can be reproduced by two dimensional molecular dynamics simulations
with good reliability, for the case of horizontal motion the results
of simulations are inconsistent with the experimental results: in {\em
  all} experimental investigations it was reported that the material
flows downwards close to the vertical
walls~\cite{TennakoonBehringer:1997,Jaeger,RosenkranzPoeschel:1996,movies},
but reported numerical simulations systematically show surface rolls
in opposite direction accompanying the more realistic deeper rolls, or
even replacing them completely~\cite{LiffmanMetcalfeCleary:1997}.

Our investigation is thus concerned with the convection pattern, i.e. the
number and direction of the convection rolls in a two dimensional
molecular dynamics simulation. We will show that the choice of the
dissipative material parameters has crucial influence on the convection pattern
and, in particular, that the type of convection rolls observed experimentally
can be 
reproduced by using sufficiently high dissipation constants.

\section{Numerical Model}
The system under consideration is sketched in Fig.~\ref{fig:sketch}:
we simulate a two-dimensional vertical cross section of a three-dimensional
container.
This rectangular section of width $L=100$ (all units in cgs system), and
infinite height, contains $N=1000$ spherical particles. The system is
periodically driven by an external oscillator $x(t) = A \sin (2\pi f
t)$ along a horizontal plane. For the effect we want to show, a
working frequency $f=10$ and amplitude $A=4$ is
selected. 
These values give an acceleration amplitude of approximately $16 g$.
Lower accelerations affect the intensity of the
convection but do not change the basic features of the convection 
pattern which we want to discuss. 
As has been shown in~\cite{SPEpre99},
past the fluidization point, a much better indicator of the convective
state is the dimensionless velocity $A 2\pi f/ \sqrt{Lg}$. This means
that in small containers motion saturates earlier, hence,  results for
different container lengths at the same values of the acceleration amplitude 
cannot be compared directly. Our acceleration amplitude $\approx 16g$ corresponds to
$\approx 3g$ in a 10 cm container (provided that the frequency is the same
and particle sizes have been 
scaled by the same amount).

The radii of the particles of density $2$ are homogeneously
distributed in the interval $[0.6, 1.4]$. The rough inner walls of the
container are simulated by attaching additional particles of the same
radii and material properties (this simulation technique is similar to ``real''
experiments, e.g.~\cite{JaegerVert}). 

For the molecular dynamics simulations, we apply a modified
soft-particle model by Cundall and Strack~\cite{CundallStrack:1979}:
Two particles $i$ and $j$, with radii $R_i$ and $R_j$ and at positions
$\vec{r}_i$ and $\vec{r}_j$, interact if their compression $\xi_{ij}=
R_i+R_j-\left|\vec{r}_i -\vec{r}_j\right|$ is positive. In this case
the colliding spheres feel the force
 $F_{ij}^{N} \vec{n}^N + F_{ij}^{S} \vec{n}^S$, 
with $\vec{n}^N$ and $\vec{n}^S$ being the unit vectors in normal and shear
direction. The normal force acting between colliding spheres reads
\begin{equation}
 F_{ij}^N = \frac{Y\sqrt{R^{\,\mbox{\it\footnotesize\it eff}}_{ij}}}{1-\nu^2} 
~\left(\frac{2}{3}\xi_{ij}^{3/2} + B \sqrt{\xi_{ij}}\, 
\frac{d {\xi_{ij}}}{dt} \right)
\label{normal}
\end{equation}
where $Y$ is the Young modulus, $\nu$ is the Poisson ratio and $B$ 
is a material constant which characterizes the dissipative
character of the material~\cite{BSHP}. 
\begin{equation}
R^{\,\mbox{\it\footnotesize\it
    eff}}_{ij} = \left(R_i R_j\right)/\left(R_i + R_j\right)  
\end{equation}
 is the
effective radius. For a strict derivation of (\ref{normal})
see~\cite{BSHP,KuwabaraKono}.

For the shear force we apply the model by Haff and Werner~\cite{HaffWerner}
\begin{equation}
F_{ij}^S = \mbox{sign}\left({v}_{ij}^{\,\mbox{\it\footnotesize\it rel}}\right) 
\min \left\{\gamma_s m_{ij}^{\,\mbox{\it\footnotesize\it eff}} 
\left|{v}_{ij}^{\,\mbox{\it\footnotesize\it rel}}\right|~,~\mu 
\left|F_{ij}^N\right| \right\} 
\label{shear}      
\end{equation}
with the effective mass $m_{ij}^{\,\mbox{\it\footnotesize\it eff}} =
\left(m_i m_j\right)/\left(m_i + m_j\right)$ and the relative velocity
at the point of contact
\begin{equation}
{v}_{ij}^{\,\mbox{\it\footnotesize\it rel}} = \left(\dot{\vec{r}}_i - 
\dot{\vec{r}}_j\right)\cdot \vec{n}^S + R_i {\Omega}_i + R_j  {\Omega}_j ~.
\end{equation}
$\Omega_i$ and $\Omega_j$ are the angular velocities of the particles.
 
The resulting momenta $M_i$ and $M_j$ acting upon the particles are
$M_i = F_{ij}^S R_i$ and $M_j = - F_{ij}^S R_j$. Eq.~(\ref{shear})
takes into account that the particles slide upon each other for the
case that the Coulomb condition $\mu \mid F_{ij}^N \mid~<~\left| 
F_{ij}^S \right|$ holds, otherwise they feel some viscous friction.
By means of $\gamma _{n} \equiv BY/(1-\nu ^2)$ and $\gamma _{s}$,
normal and shear damping coefficients, energy loss during particle
contact is taken into account~\cite{restitution}.

The equations of motion for translation and rotation have been solved
using a Gear predictor-corrector scheme of sixth order
(e.g.~\cite{AllenTildesley:1987}).

The values of the coefficients used in simulations are $Y/(1-\nu
^2)=1\times 10^{8}$, $\gamma _{s}=1\times 10^{3}$, $ \mu =0.5$. For
the effect we want to show, the coefficient $\gamma _{n}$ takes values within the range
$\left[10^2,10^4\right]$.

\section{Results}
\begin{figure}
\centerline{\psfig{file=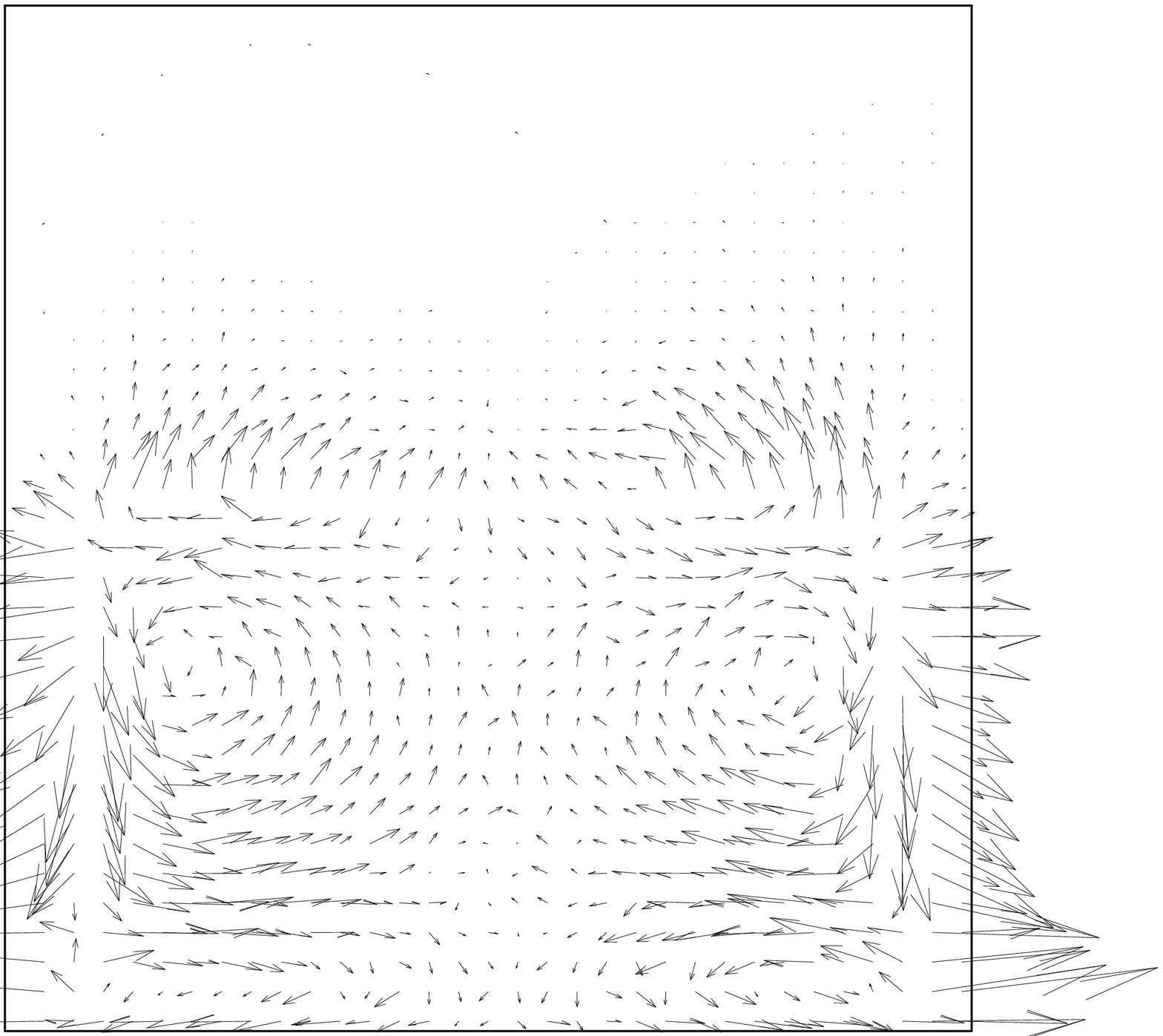,width=5.7cm,clip=}
\hspace{0.3cm}\psfig{file=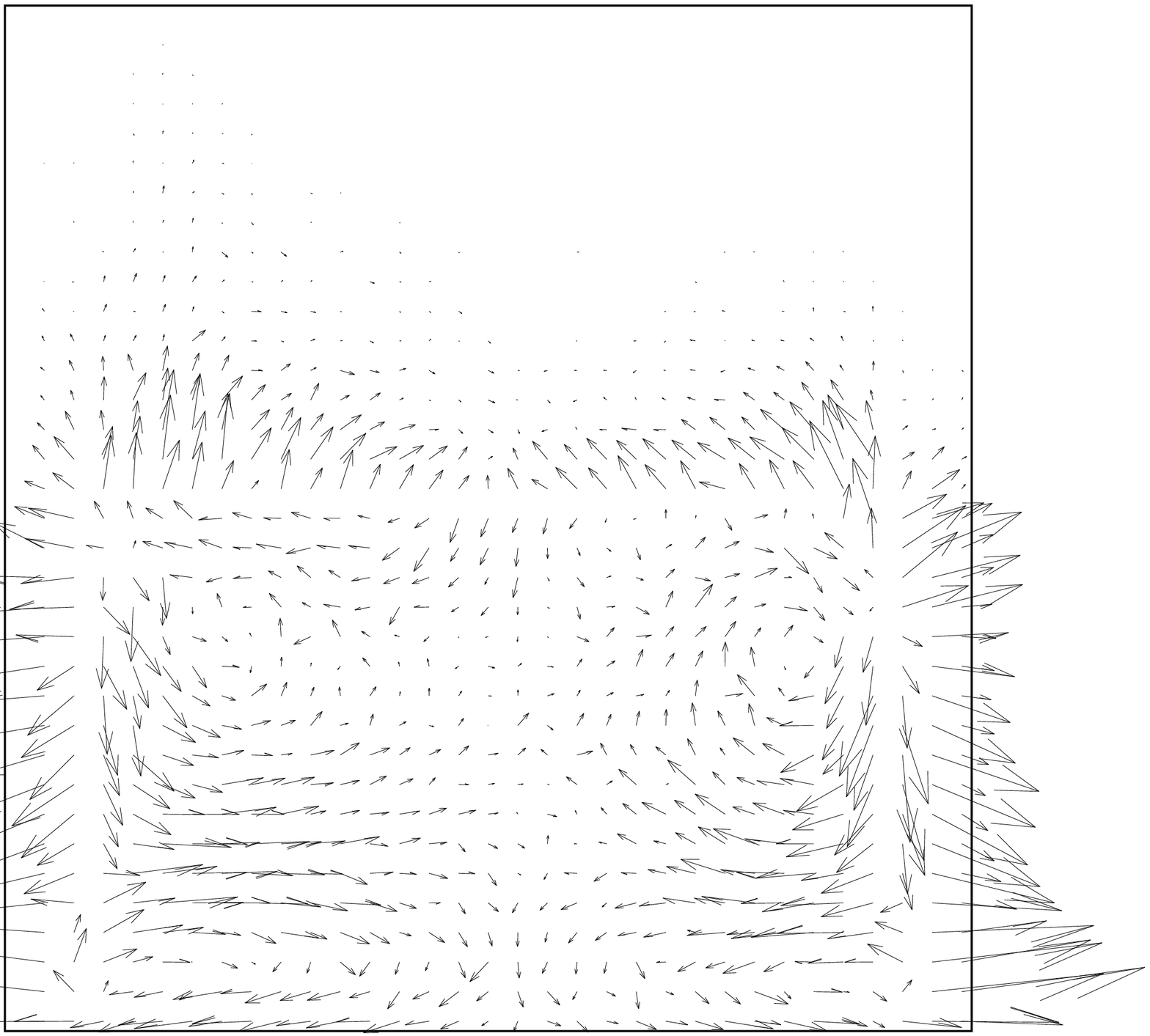,width=5.7cm,clip=}
\hspace{0.3cm}\psfig{file=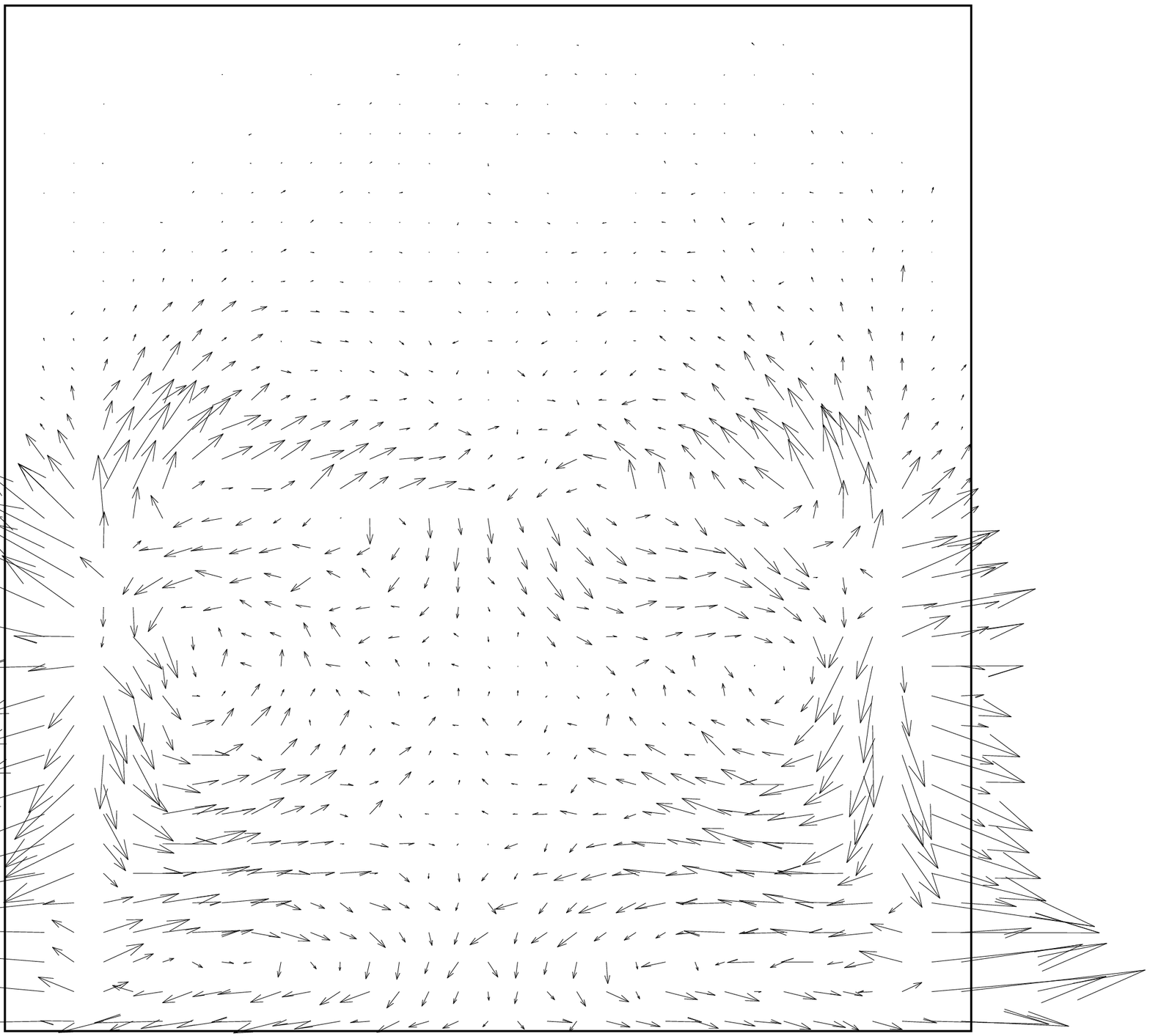,width=5.7cm,clip=}}
\centerline{\psfig{file=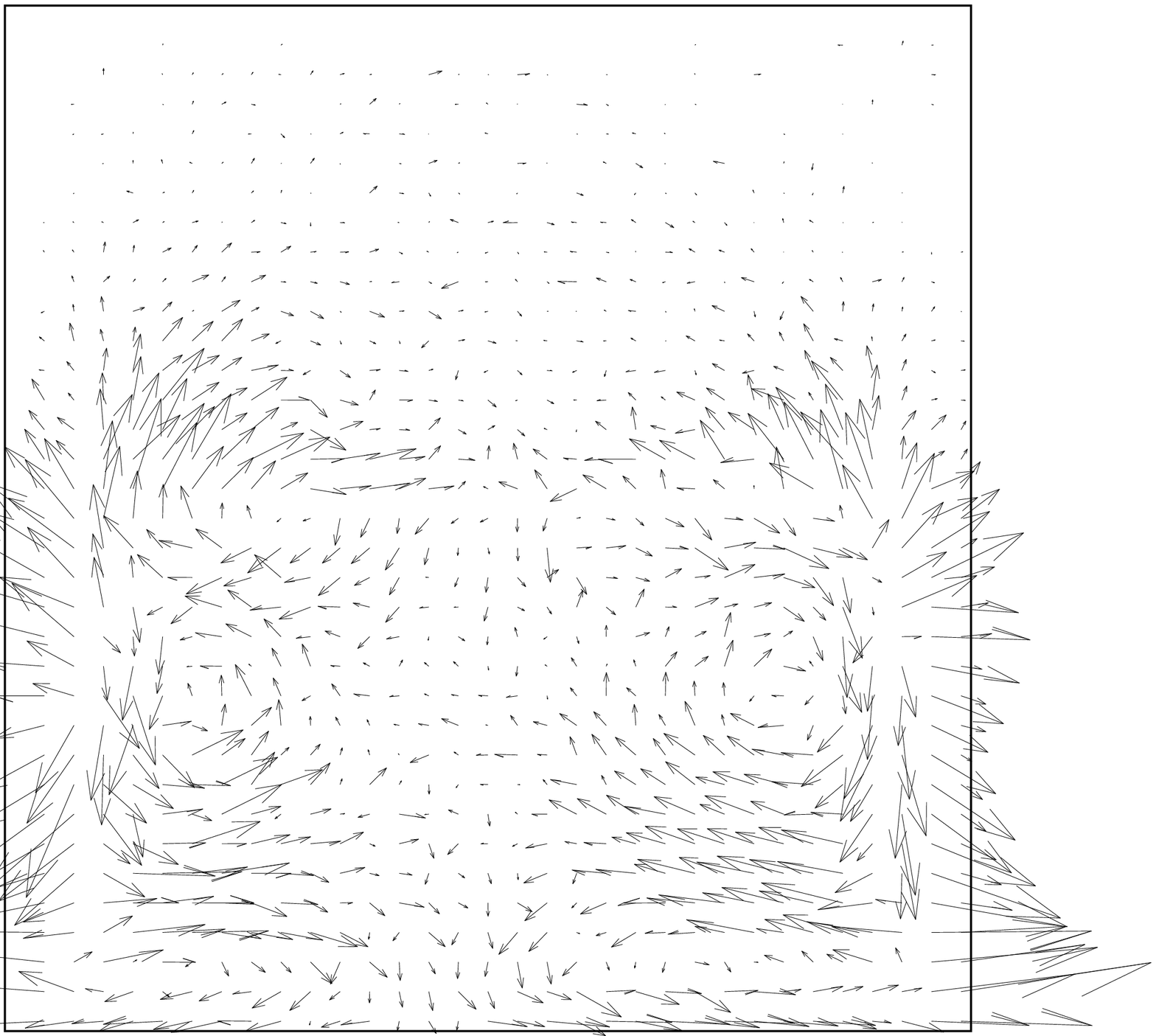,width=5.7cm,clip=}
\hspace{0.3cm}\psfig{file=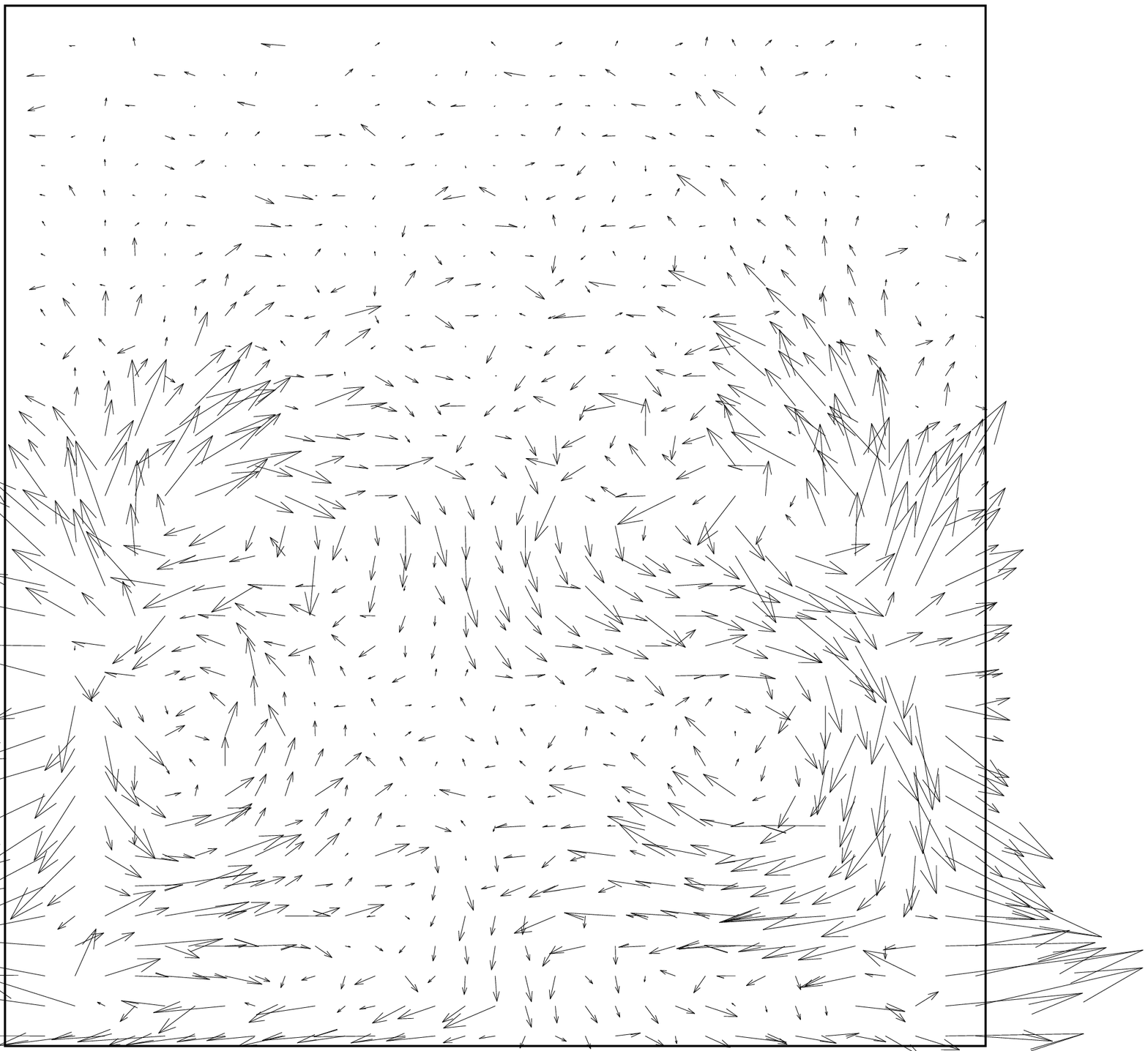,width=5.7cm,clip=}
\hspace{0.3cm}\psfig{file=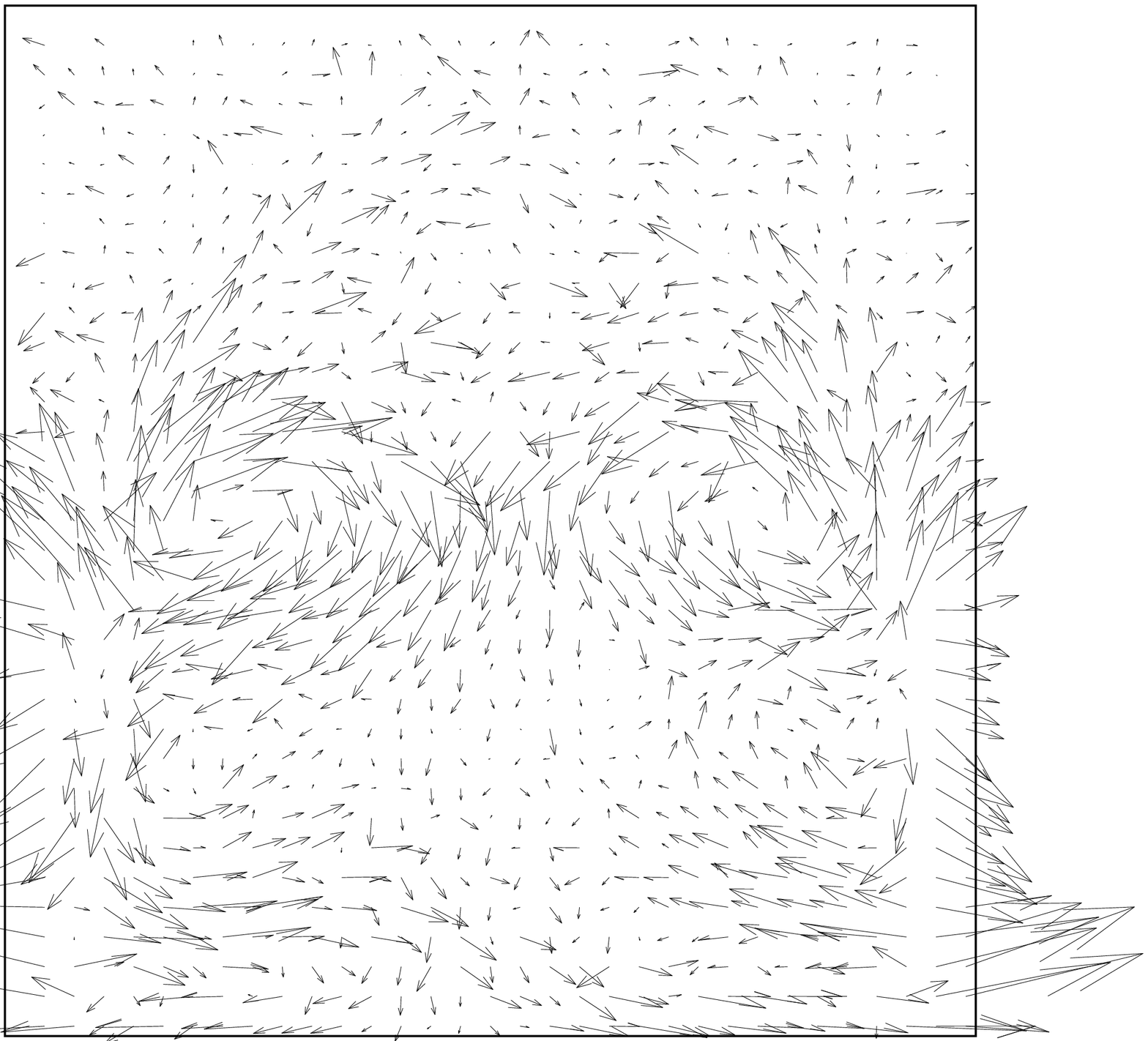,width=5.7cm,clip=}}
\centerline{\psfig{file=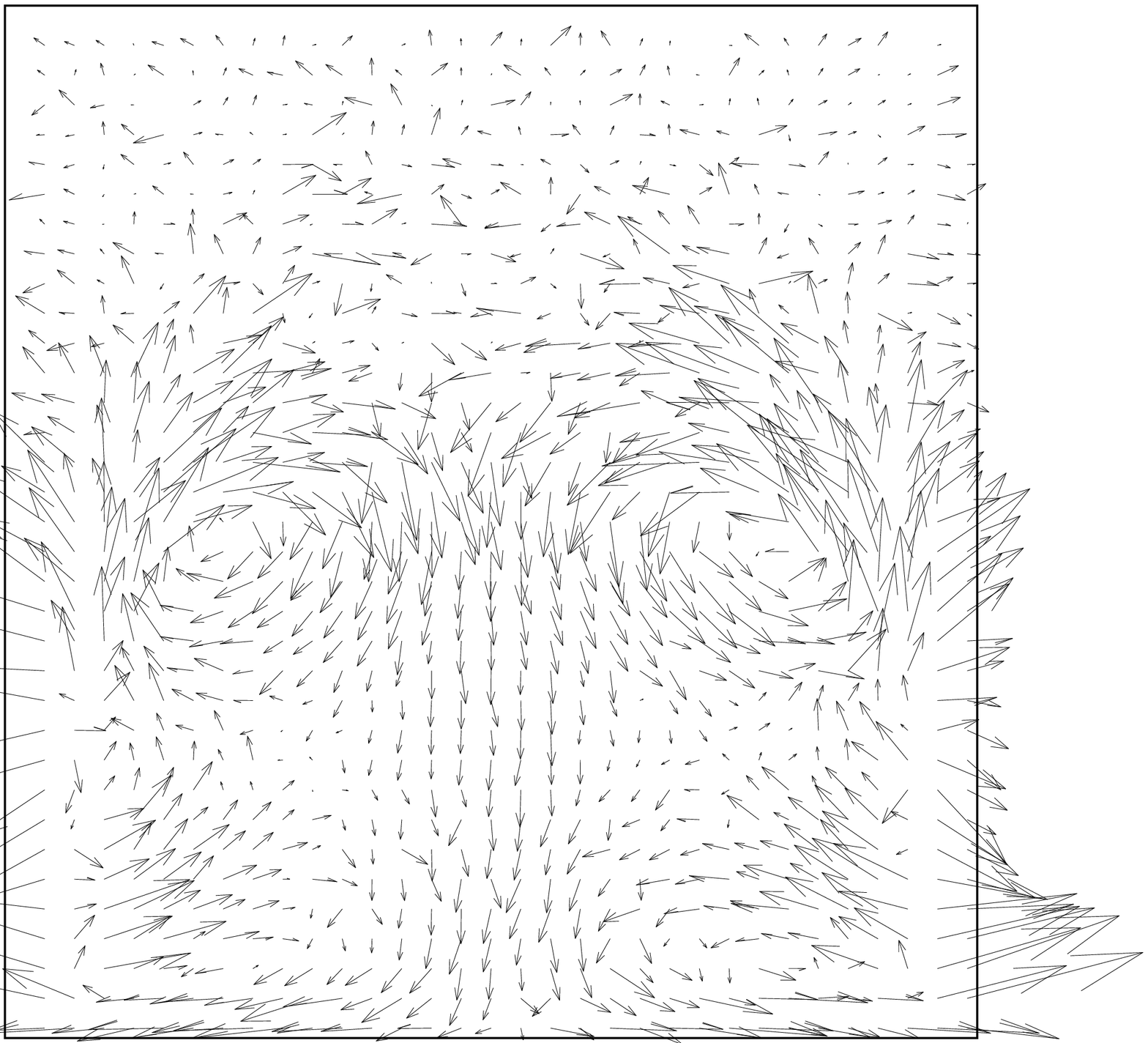,width=5.7cm,clip=}
\hspace{0.3cm}\psfig{file=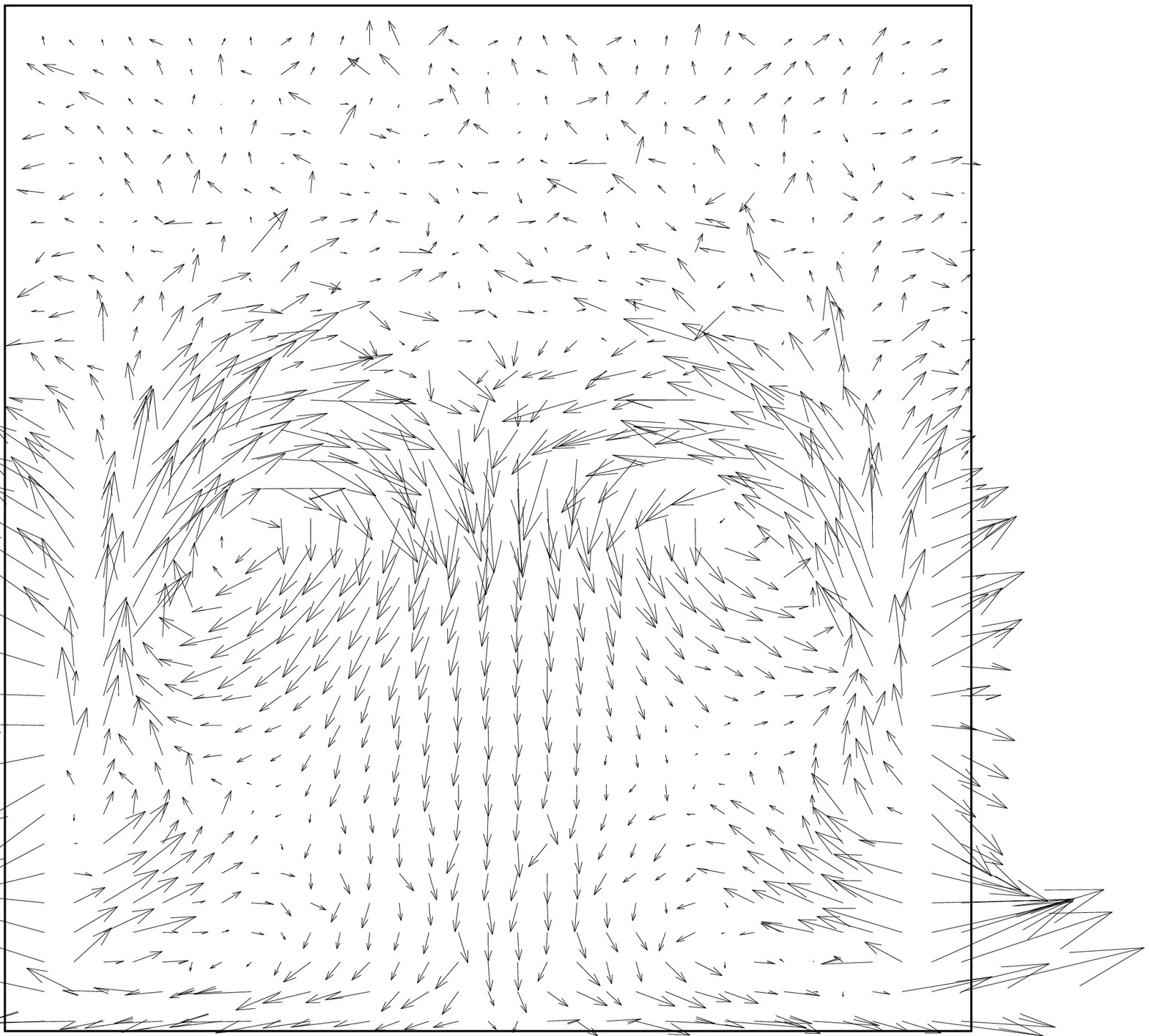,width=5.7cm,clip=}
\hspace{0.3cm}\psfig{file=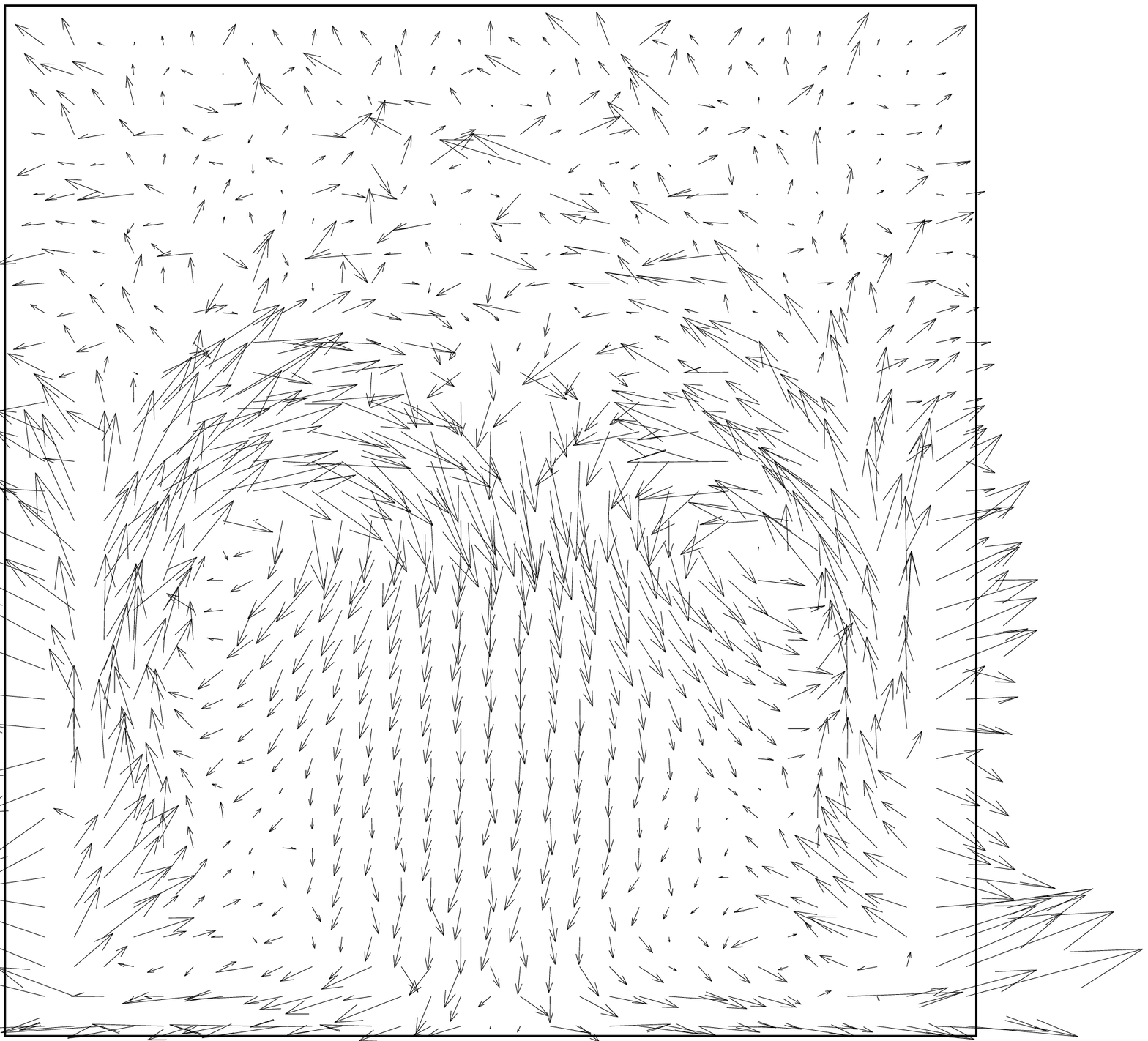,width=5.7cm,clip=}}
\vspace{0.3cm}
\caption{Velocity field obtained after cycle averaging of 
  particle displacements, for different values of the normal damping
  coefficient, $\gamma_n$. The first one is $1\times 10^4$, and for
  obtaining each subsequent frame the coefficient has been divided by
  two. The frames are ordered from left to right and from top to
  bottom. The cell size for averaging is approximately one particle diameter.}
\label{fig1}
\vspace*{-0.2cm}
\end{figure}

The mechanisms for convection under horizontal shaking have been
discussed in \cite{LiffmanMetcalfeCleary:1997}. Now we can show that
these mechanisms can be better understood by taking into account the
particular role of dissipation in this problem. The most striking
consequence of varying the normal damping coefficient is the change
in organization of the convective pattern, i.e. the direction and
number of rolls in the stationary regime. This is shown in
Fig.~\ref{fig1}, which has been obtained after averaging particle
displacements over 200 cycles 
(2 snapshots per cycle).
The asymmetry of compression and expansion of particles close to
the walls (where the material results highly compressible) explains 
the large transverse velocities shown in the figure.
Note, however, that the upward and downward motion at the walls cannot be altered 
by this particular averaging procedure. 

The first frame shows a convection pattern with only two rolls, where
the arrows indicate that the grains slide down the walls, with at most
a slight expansion of the material at the surface. 
There are no surface rolls.
This is very
similar to what has been observed in
experiments\cite{TennakoonBehringer:1997,Jaeger,RosenkranzPoeschel:1996}.
In this case, dissipation is high enough to damp most of the sloshing
induced by the vertical walls, and not even the grains just below the
surface can overcome the pressure gradient directed downwards.

For lower damping, we see the developing of surface rolls, 
which
coexist with the inner rolls circulating in the opposite way. Some
energy is now available for upward motion when the walls compress the
material fluidized during the opening of the wall ``gap'' (empty space
which is created alternatively during the shaking motion). This is the
case reported in \cite{LiffmanMetcalfeCleary:1997}. The last frames
demonstrate how the original rolls vanish at the same time that the
surface rolls grow occupying a significant part of the system.
Another feature shown in the figure is the thin layer of material involving
3 particle rows close to the bottom, which perform a different kind
of motion. This effect, which can be seen in all frames,
is due to the presence of the constraining boundaries
but has not been analyzed separately.
\onecolumn
\twocolumn

\begin{figure}
  \centerline{\psfig{file=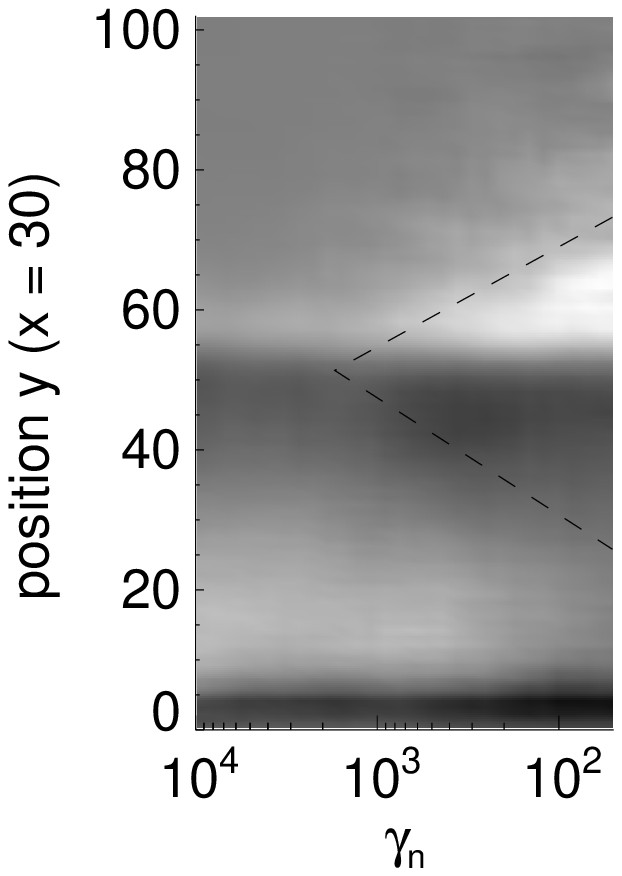,width=4.5cm,clip=}\hspace{-0.5cm}
    \psfig{file=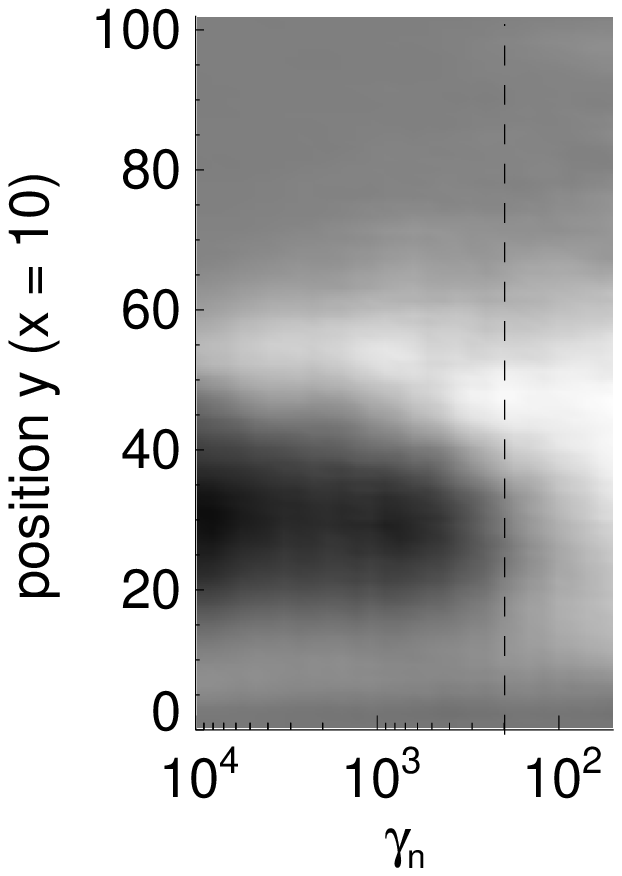,width=4.5cm,clip=}
}
\centerline{\psfig{file=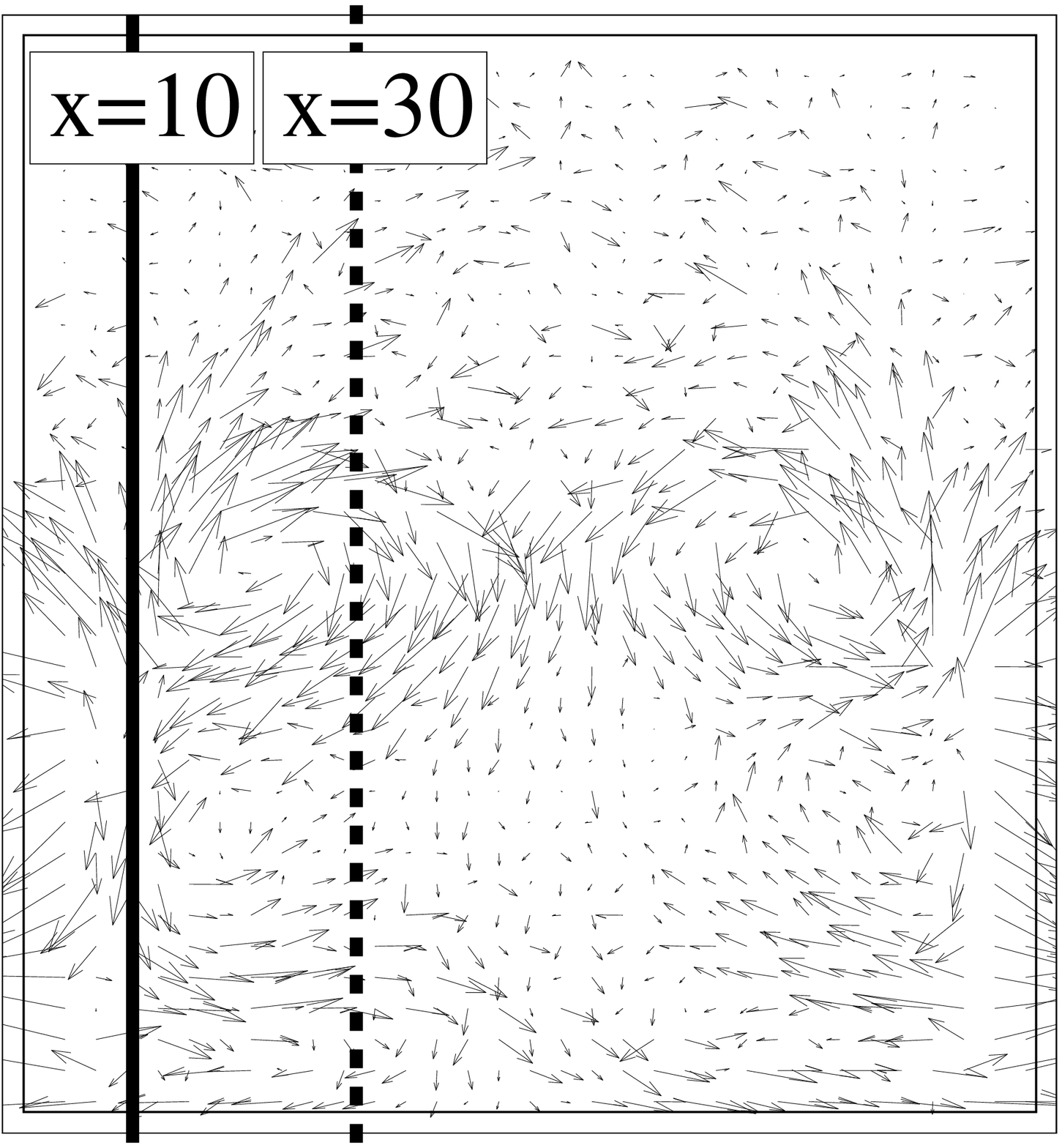,height=4.2cm,bbllx=7pt,bblly=16pt,bburx=507pt,bbury=544pt,clip=}}
\vspace*{0.2cm}
\caption{Horizontal (left) and vertical (right) displacements at 
  selected positions of the frames in Fig.~\ref{fig1} (see the text
  for details), for decreasing normal damping and as a function of
  depth. White indicates strongest flow along positive axis directions
  (up,right), and black the corresponding negative ones. The black region 
  at the bottom of the left picture corresponds to the complex boundary
  effect observed in Fig.~\ref{fig1}, involving only two particle layers.
  The 
  figure below shows a typical convection pattern together with the sections
  at $x=10$ and $x=30$ at which the displacements were recorded.}
\label{fig2}
\vspace*{-0.1cm}
\end{figure}

With decreasing normal damping $\gamma_n$ there are two transitions 
observable in Fig.~\ref{fig1}, meaning that the convection pattern changes
qualitatively at these two particular values of $\gamma_n$:
The first transition leads to the appearance of two surface rolls
laying on top of the bulk cells and circulating in opposite direction.
The second transition eliminates the bulk rolls. A more detailed analysis of 
the displacement fields  (Fig.~\ref{fig2})
allows us to locate the transitions much more precisely.
In Fig.~\ref{fig2} we have represented in  grey-scale the horizontal and
vertical components of the displacement vectors pictured in
Fig.~\ref{fig1} but in a denser sampling, analyzing data from 30 simulations 
corresponding to 
values of the normal damping coefficient within the interval [50,10000]. 
For horizontal displacements, we have chosen vertical sections 
at some representative position in horizontal direction
($x=30$). For the vertical displacements, vertical sections of the
leftmost part of the container were selected ($x=10$), s.
Fig.~\ref{fig2}, lower part.

The horizontal axis shows the values of the normal damping
coefficient scaled logarithmically in decreasing sequence. The
vertical axis represents the position in vertical direction, with the
free surface of the system located at $y \approx 60$.  One observes first
that white surface shades, complemented by subsurface black ones,
appear quite clearly at about $\gamma =$2000 in Fig.~\ref{fig2}
(left), indicating the appearance of surface rolls. On the other
hand, Fig.~\ref{fig2} (right) shows a black area (indicative of
downward flow along the vertical wall) that vanishes at
$\gamma_n \approx 200$ (at this point the grey shade represents vanishing vertical velocity). 
The dashed lines in Fig.~\ref{fig2} lead the eye to identify the transition values.
In the interval $ 200 \lesssim \gamma_n
\lesssim 2000$ surface and inner rolls coexist, rotating in opposite
directions.

One can analyze the situation in terms of the restitution coefficient.
\ From Eq. (\ref{normal}), the equation of motion for the displacement
$\xi_{ij}$ can be integrated and the relative energy loss in a
collision $\eta=(E_0-E)/E_0$ (with $E$ and $E_0$ being the energy of
the relative motion of the particles) can be evaluated approximately.
Up to the lowest order in the expansion parameter, one
finds~\cite{Thomas-Thorsten}
\begin{equation}
\eta = 1.78 \left( \frac{\tau}{\ell} v_0\right)^{1/5}\;,
\label{energyloss}
\end{equation}
where $v_0$ is the relative initial velocity in normal direction, and
$\tau$, $\ell$, time and length scales associated with the problem
(see~\cite{Thomas-Thorsten} for details),

\begin{equation}
\tau = \frac{3}{2} B\; ,~~~~~~~~~
\ell = \left(\frac{1}{3} \frac{m_{ij}^{\,\mbox{\it\footnotesize\it eff}} 
}{\sqrt{R^{\,\mbox{\it\footnotesize\it eff}}_{ij}} 
B \gamma_{n}}\right)^{2}.
\end{equation}
For $\gamma_n = 10^4$ (the highest value analyzed) and the values of
the parameters specified above ($v_0 \approx A 2\pi f$ for collisions
with the incoming wall), $B= 10^{-4}$ and $\eta$ is typically
50\%. This means that after three more collisions the particle leaves
with an energy not enough to overcome the height of one single
particle in the gravity field. For $\gamma_n = 10^3$ and the other
parameters kept constant, $B=10^{-5}$ and $\eta$ has been
reduced to 5\%, resulting in that the number of collisions needed for
the particle to have its kinetic energy reduced to the same residual
fraction, has increased roughly by an order of magnitude. On the other
hand, given the weak dependence of Eq. (\ref{energyloss}) on the
velocity, one expects that the transitions shown in Fig.~\ref{fig2}
will depend also weakly on the amplitude of the shaking velocity. The reduction of the
inelasticity $\eta$ by an order of magnitude seems enough for
particles to ``climb'' the walls and develop the characteristic
surface rolls observed in numerical simulations.

\section{Discussion}
We have shown that the value of the normal damping coefficient
influences the convective pattern of horizontally shaken granular
materials. By means of molecular dynamics simulations in two
dimensions we can reproduce the pattern observed in real experiments,
which corresponds to a situation of comparatively high damping,
characterized by inelasticity parameters $\eta$ larger than 5\%. For
lower damping, the upper layers of the material develop additional
surface rolls as has been reported previously. As normal damping
decreases, the lower rolls descend and finally disappear completely at
inelasticities of the order of 1\%.

\begin{acknowledgement}
The authors want to thank R. P. Behringer, H. M. Jaeger, M. Medved,
and D. Rosenkranz for providing experimental results prior to
publication and V. Buchholtz, S. E. Esipov, and L. Schimansky-Geier
for discussion. The calculations have been done on the parallel
machine {\it KATJA} (http://summa.physik.hu-berlin.de/KATJA/) of the
medical department {\em Charit\'e} of the Humboldt University Berlin.
The work was supported by Deut\-sche Forschungsgemeinschaft through
grant Po 472/3-2.
\end{acknowledgement}


\begin{thebibliography}{99}
\bibitem{restitution} Several authors describe the dissipation of colliding particles by a ``coefficient of restitution'' $\epsilon$. Applying the correct normal force (\ref{normal}) for colliding non-cohesive particles~\cite{BSHP,KuwabaraKono} one finds that $\epsilon$ depends on their relative velocity~\cite{BSHP,Thomas-Thorsten}. Therefore, one cannot express $\gamma_n$ in terms of $\epsilon$.
\end{thebibliography}
\end{document}